\documentclass[mathleft
]{an}
\usepackage{graphicx}
\usepackage{times}
\begin{document}

\Pagespan{0}{}
\Yearpublication{}%
\Yearsubmission{}%
\Month{}%
\Volume{}%
\Issue{}%

\title{How environment drives galaxy evolution: lessons learnt from satellite galaxies}

\author{A. Pasquali\inst{1}\fnmsep\thanks{Corresponding author:
  \email{pasquali@ari.uni-heidelberg.de}\newline}
}
\titlerunning{Galaxy Environment}
\authorrunning{A. Pasquali}
\institute{
Astronomisches Rechen-Institut, Zentrum f\"ur Astronomie der Universit\"at Heidelberg,
M\"onchhofstrasse 12 - 14, D-69120 Heidelberg, Germany
}



\abstract{It is by now well established that galaxy evolution is driven by intrinsic and environmental
processes, both contributing to shape the observed properties of galaxies. A number of early
studies, both observational and theoretical, have shown that the star formation activity of galaxies 
depends on their environmental local density and also on galaxy hierarchy, i.e. centrals vs. satellites.
In fact, contrary to their central (most massive)
galaxy of a group/cluster, satellite galaxies are stripped of their gas and stars, and
have their star formation quenched by their environment. Large galaxy surveys like SDSS now
permit us to investigate in detail environment-driven transformation processes by comparing
centrals and satellites. In this paper I summarize what we have so far learnt about
environmental effects by analysing the observed properties of local central and satellite galaxies
in SDSS, as a function of their stellar mass and the dark matter mass of their
host group/cluster.}

\maketitle

\section{Introduction}
As shown by the Sloan Digital Sky Survey (SDSS) which has extensively mapped the large-scale structure 
of the local Universe in space and redshift, most galaxies preferentially live in aggregates whose size and mass
range from galaxy pairs to groups and clusters. In the last three decades it has become clear that 
galaxy mass and environment together concur to shape galaxy properties. We well know, for example, that the
more massive galaxies residing in dense regions tend to have an early-type morphology as well as older and
metal-richer stellar populations, and to exhibit a rather low star-formation activity (aka the density - morphology 
relation, Dressler 1980). On the contrary, galaxies experiencing a significant rate of star formation are
typically found in low-density environments (Hashimoto et al. 1998). At present, we still need to understand in details 
the physics of environmental effects and their interplay with the internal secular evolution of galaxies.
\par
An obstacle to this kind of studies comes from the definition of environment itself. Environment has usually been
measured in terms of galaxy density (i.e. the number density of galaxies out to the $n$-th neighbour), although
such a measurement in a rich galaxy group or cluster is representative only of the local environment and not of
the whole group/cluster. The advent of large surveys has made it possible to use group/cluster finding algorithms,
and thus quantify environment on a physical basis, in terms of its encompassing dark matter halo, whose mass is estimated 
from the stellar mass of the galaxies in each group/cluster (cf. Yang et al. 2005, 2007) and links environment to the 
evolution of structure in the Universe. Within each environment it becomes also possible to distingusih galaxies 
between the central (the most massive) galaxy and the satellites. These two classes of galaxies 
are in fact foreseen by simulations and semi-analytic models (SAMs) of galaxy formation and evolution to follow different
assembly and star-formation histories. The distinction between centrals and satellites allows us to directly compare
their observed properties with those predicted by simulations and SAMs in order to better constrain the mechanisms
through which environment affects galaxy evolution.
\par
Contrary to their central galaxy, satellites orbit within their environment and interact with the local Intra-Cluster
Medium (ICM), their peers and the potential well of their group/cluster. These interactions lead to erosion processes
which cause satellites to lose their reservoir of hot gas (aka strangulation, cf. Larson et al. 1980),
ionized and neutral gas (ram-pressure stripping, Gunn \& Gott 1972), as well as their stars via tidal stripping (e.g. Kang 
\& van den Bosch 2008, Pasquali et al. 2010).
Harassment, due to fast encounters (Moore et al. 1998), and tidal interactions with their central galaxies can also remove gas
and stars from satellites. Our goal is to use the observed properties of satellites to quantify the amplitude and time scale
of environmental effects, and how these parameters depend on galaxy stellar mass and group/cluster halo mass, on cluster-centric distance and redshift. Therefore, we will review what has been learnt on
galaxy environment at redshift $z \simeq$ 0 in Sect. 2 and 3, and we will describe the current 
knowledge of environmental effects at higher redshifts in Sect. 4. Conclusions will follow in Sect. 5.

\section{Quenching star formation}
Since galaxy evolution is driven by galaxy stellar mass (M$_{\star}$) and environment at the same time, we need to compare the 
observed properties of satellites at fixed M$_{\star}$ in order to characterize environmental effects alone. Moreover, in the assumption
that satellites progenitors have similar properties to present-day centrals of the same M$_{\star}$, the comparison between
satellites and centrals in their properties at fixed M$_{\star}$ allows us to further constrain those galaxy transformations that
are induced solely by environment. 

In terms of their specific star formation rate (SSFR = SFR/M$_{\star}$), most of the satellites residing in massive haloes
at $z \sim$ 0 are passive, i.e. they are characterized by log(SSFR/yr$^{-1}) <$ -11. In particular, at fixed M$_{\star}$
the fraction of passive satellites increases with halo mass (M$_{\rm h}$) and is higher than the fraction of equally-massive passive
centrals at least for log(M$_{\star}$/M$_{\odot}$) $<$ 10.9 (Wetzel et al. 2012, also van den Bosch et al. 2008). Moreover, 
the fraction of passive satellites  is seen to steadily increase with M$_{\star}$ in any environment of given halo mass. This 
indicates that star formation is most efficiently suppressed in satellites by both their intrinsic evolution (i.e. secular evolution 
and/or AGN
feedback especially at high M$_{\star}$) and environment (an effect that becomes more significant at low M$_{\star}$ and 
high M$_{\rm h}$).
This trend does not imply, though, that all satellites in massive haloes are quenched; their distribution in SSFR is bimodal irrespective of M$_{\rm h}$, only the amplitude of the peaks of the star-forming and quenched satellites changes with M$_{\star}$
and M$_{\rm h}$. At fixed M$_{\rm h}$, the fraction of passive satellites increases with M$_{\star}$ as more satellites quench their star formation activity through their secular evolution or AGN feedback. At fixed M$_{\star}$, the fraction of passive satellites increases as more star-forming satellites are quenched by environment with increasing M$_{\rm h}$. The transition region between the  peaks of the star-forming and passive satellites is, though, invariant with respect to M$_{\rm h}$, being narrow in log(SSFR) and always occurring at log(SSFR/yr$^{-1}$) $\simeq$ -11 (Wetzel et al. 2012). Such a finding may imply that star formation in satellites is quenched on a short time scale, independent of M$_{\rm h}$. 
\par

\begin{figure}
\includegraphics[width=7cm]{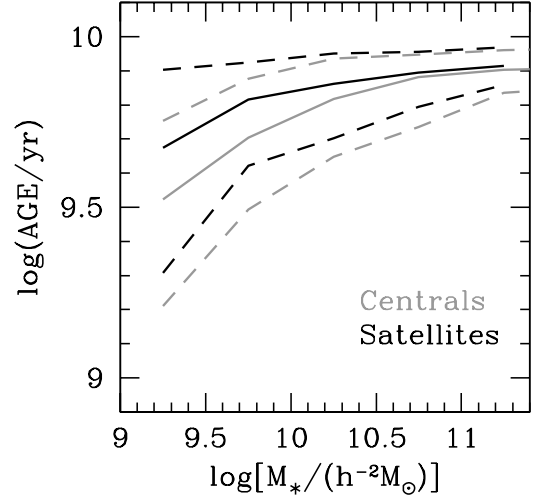}
\caption{The average stellar age (weighted by mass) - M$_{\star}$ relation (solid lines) for centrals (in grey) and satellites 
(in black). The dashed lines represent the 16th and 84th percentiles of the stellar age distribution within each bin of stellar mass (Pasquali et al. 2010).}
\end{figure}

A rapid quenching is also consistent with the fact that low-mass satellites are older
than equally-massive centrals, typically by just 1.5 Gyr at log[M$_{\star}$/(h$^{-2}$M$_{\odot}$)] $\simeq$ 9.5. Figure 1 shows the average stellar age (weighted by mass) - M$_{\star}$ relation (solid lines) for centrals (in grey) and satellites (in black). 
It can be seen that satellites less
massive than log[M$_{\star}$/(h$^{-2}$M$_{\odot})] \simeq$ 10.5 are older than centrals of the same stellar mass, and this age difference 
becomes larger with decreasing M$_{\star}$. At fixed M$_{\star}$, the average stellar age of satellites tends to increase with 
M$_{\rm h}$, and this trend is more significant for satellites with log[M$_{\star}$/(h$^{-2}$M$_{\odot})] <$ 10. 
This can be explained if low-mass satellites in more massive haloes had an earlier epoch of accretion with respect to low-mass
satellites in less massive environments (Pasquali et al. 2010). In this framework, they ceased their star formation activity
early on, and their stars evolved passively till now. 
\par
The observed properties described above were used by Wetzel et al. (2013, see also Bah\'e et al. 2015) to 
depict the ``delayed-then-rapid'' quenching scenario
for satellites, where low-mass galaxies are accreted onto a bigger halo at an earlier redshift (with respect to more massive galaxies),
and keep forming stars and consuming their cold gas at the same rate for  2 - 4 Gyr after infall. Meanwhile, 
tidal forces induced by their
group/cluster potential well deprive satellites of their hot gas (strangulation), i.e. their gas reservoir for future star formation. 
After this period of time, satellites extinguish their star formation rapidly, with an e-folding time $\leq$ 0.8 Gyr.

\subsection{Ram-pressure stripping}
A mechanism able to induce a rapid quenching of star formation may be identified in ram pressure. The interaction between
an orbiting satellite and the ICM of its host halo triggers gas losses from the satellite, and the fraction of stripped gas
is directly proportional to the satellite square orbital velocity, the ICM density and the group/cluster M$_{\rm h}$ (Bekki 2009,
Kapferer et al. 2009). Satellites subjected to ram-pressure stripping are well known to exhibit a bent and distorted spatial 
distribution of HI and H$\alpha$, together with a gaseous wake possibly undergoing star formation (e.g. Kenney et al. 
2004, 2014, Vollmer et al. 2004).
\par
There is more indirect evidence for the occurrence of ram-pressure stripping encoded in the observed properties of satellites.
This is the case when, for example, the fraction of passive satellites is computed as a function of cluster-centric distance. 
The probability for a satellite of given M$_{\star}$ to be quenched increases with decreasing distance from the group/cluster
centre, and such a trend is more pronounced for low-mass satellites which were likely accreted at an earlier time. This observational
result can be explained in terms of the ICM density being larger in the central region of a group/cluster than in its outskirts.
In addition, the orbital velocity of satellites at pericenter is significantly higher than at apocenter.
Ram-pressure stripping is thus expected to be more efficient for satellites located in the inner parts of their host halo.
\par\noindent
Furthermore, radial profiles of the fraction of passive satellites show the existence of quenched satellites also at
distances larger than the halo virial radius. These satellites may be backsplash galaxies moving on an eccentric orbit
which took them across the central region of their host halo (thus exposing them to ram-pressure and tidal stripping), and has now
brought them outside the halo virial radius. Alternatively, the accretion of a smaller galaxy group with its own quenched satellites as well as the infalling of galaxies onto an extended ellipsoidal (rather than spherical) halo can also increase the 
fraction of quenched satellites at R $>$ R$_{\rm vir}$ (Wetzel et al. 2012, Bah\'e et al. 2013, Muriel \& Coenda 2014).
\par
Ram pressure is likely to be more efficient when acting on gas more weakly bound to a satellite, as in the case of the HI gas, generally
distributed on a wider spatial scale with respect, for example, to the ionized H$\alpha$ gas used to estimate star formation rates. 
Indeed, Fabello et al. (2012) have shown that the fraction of galaxies with a measured HI mass (from the ALFALFA survey, Giovanelli
et al. 2005) declines 
much more rapidly than the fraction of star-forming galaxies in haloes more massive than log(M$_{\rm h}$/M$_{\odot}$) = 13.
Moreover, Gavazzi et al. (2013) have found that the HI content of satellites in the Local Supercluster significantly diminuishes
at small distances from M~87 where the ICM density is likely the highest.
\par
The removal of cold and ionized gas clearly reduces the galaxy star formation rate; when compared at fixed M$_{\star}$, satellites
tend to exhibit a lower global SSFR (as derived from their integrated colours, Brinchmann et al. 2004) than centrals (Pasquali et al.
2012). In addition, the stripping of metal-poor gas from the galaxy outskirts may have an impact also on the galaxy gas-phase metallicity. In Figure 2 satellites (in black) are compared with equally massive centrals (in grey) on the basis of their gas-phase
metallicity measured in the SDSS fibre by Tremonti et al. (2004). Therefore, the gas-phase metallicity shown in Fig. 2 is the
oxygen metallicity of the Interstellar Medium in the central region of galaxies. It can be seen that satellites tend to be metal-richer
than centrals of the same M$_{\star}$ and the difference in 12 $+$ log(O/H) increases with decreasing M$_{\star}$ so to reach a
maximum value of 0.06 dex at log[M$_{\star}$/(h$^{-2}$M$_{\odot})] \simeq$ 8.2 (Pasquali et al. 2012). The gas-phase metallicity of satellites less massive than log[M$_{\star}$/(h$^{-2}$M$_{\odot})] \simeq$ 9.5 becomes higher with increasing M$_{\rm h}$; in particular,
at log[M$_{\star}$/(h$^{-2}$M$_{\odot})] \simeq$ 9 it increases by 0.15 dex in the range 11 $<$ log[M$_{\rm h}$/(h$^{-1}$M$_{\odot})] <$ 14.
These findings are consistent with ram pressure which, by stripping the outermost gas, is able to inhibit radial inflows of metal-poor 
gas that would otherwise diluite the gas-phase metallicity in the inner region of satellites. Since 
low-mass satellites in present-day, more masive haloes were accreted earlier, they have been
exposed to ram-pressure stripping for longer, and have been able to preserve their
gas-phase metallicity at the time of infall better than their peers in present-day, less massive
haloes (Pasquali et al. 2012). 

\begin{figure}
\includegraphics[width=7cm]{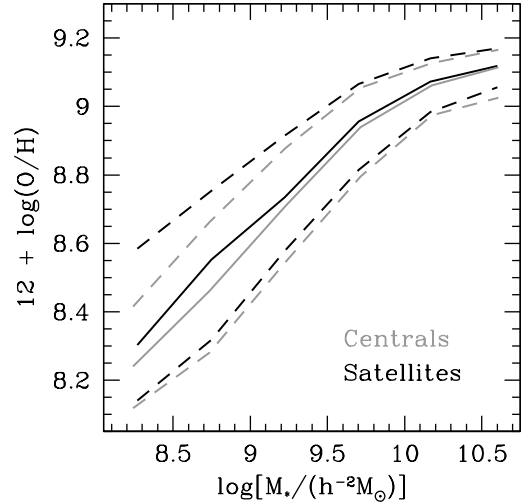}
\caption{The average gas-phase metallicity of centrals (grey solid line) and satellites (black solid line) as a function of 
their stellar mass. The dashed lines represent the 16th and 84th percentiles of the gas-phase metallicity distribution within each bin of stellar mass (Pasquali et al. 2012).}
\end{figure}

\subsection{Comparison with SAMs predictions and simulations}
In their original version, semi-analytic models of galaxy formation and evolution assumed that, upon accretion,
satellites would instantaneously lose their reservoir of hot gas, thus experiencing an early and rapid quenching of their 
star formation activity.
Consequently, SAMs predicted fractions of passive satellites much higher than observed, especially at low stellar masses 
(log(M$_{\star}$/M$_{\odot}) <$ 10, cf. for example Weinmann et al. 2006). An additional effect of a too early and rapid quenching
is that SAMs overestimate the stellar age of low-mass satellites whose observed stellar ages are instead younger (Pasquali et al. 2010).
\par
Recently, Guo et al. (2011) modified their SAMs by introducing a dependence of strangulation on M$_{\rm h}$; also in this case the predicted fraction of quenched satellites turns out to be larger than observed particularly at low stellar masses. The comparison between observed and predicted fractions of quenched galaxies implies, once again, that low-mass satellites residing in haloes more massive than log(M$_{\rm h}$/M$_{\odot}) \simeq$ 13 cease their star formation activity over a long time scale, of the order of 5 Gyr. Quenching should become shorter with increasing M$_{\star}$ (De Lucia et al. 2012, Hirschmann et al. 2014).
\par
An alternative way to reconcile predicted and observed fractions of passive satellites was proposed by Kang \& van den Bosch (2008),
who adopted a strangulation time scale of 3 Gyr in their SAMs and assumed that 50$\%$ of low-mass satellites 
(log(M$_{\star}$/M$_{\odot} <$ 10) are tidally
disrupted before being accreted onto their central galaxy. As discussed below, there exists a 
growing observational evidence for satellite galaxies being tidally stripped of their stars, which
may be a prelude to tidal disruption at least for some of them. The fractions of quenched satellites computed by Kang \& van den 
Bosch (2008) follow the observed ones quite closely.
\par 
The hydrodynamical simulations based on GADGET-2 and performed by Dav\'e et al. (2011, 2013) automatically include ram-pressure stripping,
and predict a gas-phase metallicity difference between satellites and centrals as well as a fraction of HI-poor satellites in good agreement with observations.

\section{Tidal stripping and stellar mass loss}
Recent deep-imaging observations at optical wavelengths (with a limiting surface brightness
$\mu_g \simeq$ 28 - 30 mag arcsec$^{-2}$) have revealed the presence of tidal 
tails, streams and shells around galaxies, of both early and late type, in low-to-intermediate 
density environments as well as in clusters. These fine structures are commonly interpreted as the 
result of recent close interactions or mergers, during which one or more companions were firstly 
tidally stripped of their stars
and later accreted (Mart\'inez-Delgado et al. 2010, Sheen et al. 2012, Duc et al. 2015).  
\par
$N$-body numerical simulations have indeed shown the emergence of these tidal features during
the assembly history of galaxies and how the morphology of such substructures depends on
the orbit type of the accreted companions (cf. for example Johnston et al. 2008). In particular,
Chang et al. (2013) simulated the gravitational interactions between a central galaxy and
its orbiting satellite in order to study the morphological transformation of the satellite galaxy
as induced by tidal stripping. These simulations indicate that tidal stripping, hence stellar
mass loss from the satellite, begins when $\sim$90$\%$ of the satellite dark matter halo has been
removed. At this point, the efficiency of tidal stripping and the amplitude of the stellar mass
loss critically depend on the satellite morphology. In the case of a pure disc or a disc$+$bulge
satellite with a tidal radius of the order of the disc scale-length, stellar mass loss is
exponential and the removal of most of the disc component occurs on a time scale of few Gyrs.
The orientation of the satellite orbit and the morphology of the central galaxy are found to
have only a second-order effect on tidal stripping.
\par
Tidal stripping and morphological transformations of satellites are also driven by the global
tidal field of their host halo. The numerical simulations by Villalobos 
et al. 
(2012) have shown that satellites can develop tidal features and undergo relevant changes in their
original morphology once they cross a region of their host group where the mean group density is 
comparable to their internal, central density. The efficiency of tidal stripping and the time scale 
over which most of the satellite stars are removed are found to depend on the satellite stellar
mass, orbit type and the satellite inclination with respect to the orbit.

\begin{figure}
\includegraphics[width=7cm]{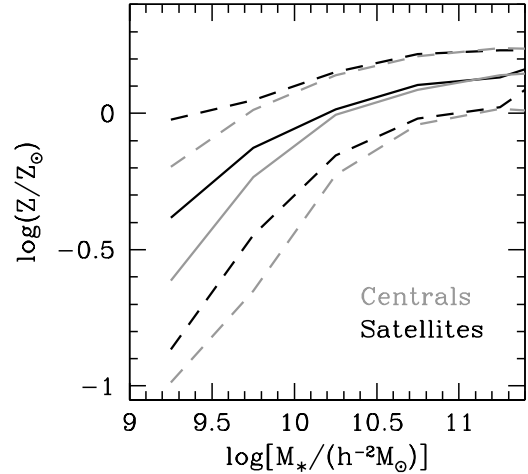}
\caption{The average stellar metallicity of 
centrals (grey solid line) and satellites (black solid line) as a function of their stellar mass. 
The dashed lines represent the 16th and 84th percentiles of the stellar metallicity 
distribution within each bin of stellar mass (Pasquali et al. 2010).}
\end{figure}

Another effect of tidal stripping, in addition to tidal features and morphological transformations,
may be identified in the observed offset in stellar metallicity between central and satellite
galaxies at fixed stellar mass. Figure 3 shows the dependence of stellar metallicity on M$_{\star}$
for centrals (solid grey line) and satellites (solid black line). While at the high-mass end central
and satellite galaxies are equally metal-rich, satellites less massive than log[M$_{\star}$/(h$^{-2}$M$_{\odot})] 
\simeq$ 10.3 exhibit a higher stellar metallicity than centrals of the same 
M$_{\star}$. This difference becomes larger for decreasing stellar mass, and reaches a values of
$\sim$0.2 dex at log[M$_{\star}$/(h$^{-2}$M$_{\odot})] \simeq$ 9.3 (Pasquali et al. 2010). Observations
also indicate the existence of a stellar metallicity - M$_{\rm h}$ relation for satellites, whereby
the stellar metallicity of low-mass satellites (log[M$_{\star}$/(h$^{-2}$M$_{\odot})] \simeq$ 10.3)
is seen increasing in more massive galaxy groups. A qualitative explanation of these
trends rests on the fact that, upon accretion, satellites lose stellar mass via tidal stripping which
leaves, though, stellar metallicity unaltered. Thus, present-day satellites are most likely less massive 
than, but equally metal-rich as, their progenitors. Comparing present-day satellites with centrals
of the same M$_{\star}$ equals comparing present-day centrals with past, more massive centrals
which are also metal-richer by virtue of the galaxy stellar metallicity - M$_{\star}$ relation.
This ultimately produces an offset in stellar metallicity when present-day satellites and
centrals are contrasted at fixed present-day M$_{\star}$. In this framework, a difference of $\sim$0.2 
dex in log(Z/Z$_{\odot}$) at log[M$_{\star}$/(h$^{-2}$M$_{\odot})] \simeq$ 9.3 would indicate that a
satellite of this present-day M$_{\star}$ has experienced a fractional mass loss of about 50$\%$. 
Since satellite galaxies in present-day, more massive environments were accreted at an earlier time, 
they have undergone the effects of tidal stripping for a longer period and, consequently, they are
expected to be metal-richer than satellites of the same M$_{\star}$ in less massive haloes (Pasquali
et al. 2010).

\section{Beyond z = 0}
We observe environmental effects also at higher redshifts, at least up to $z \simeq$ 2, although 
larger distances progressively hamper our capability of measuring detailed stellar parameters,
and limit our measurements to more massive galaxies and environments.
Most measurements are restricted to star formation rates from emission lines and photometry,
and to stellar ages from the D4000 break.
\par
A number of studies have shown that, at intermediate redshift (0.2 $< z <$ 0.8), star-forming galaxies
in groups and clusters exhibit star formation rates that are, at fixed M$_{\star}$, a factor of
two lower than those measured in the field population (cf. for example Poggianti et al. 2006, Vulcani
et al. 2010). Consequently, the fraction of passive galaxies is observed to rise, at fixed M$_{\star}$,
with halo mass, from the field population to galaxy groups and clusters. Following the time evolution
of the cosmic star formation rate, the fraction of passive satellites in groups and clusters tend
to become lower with increasing redshift and closer to the value measured in the field. At fixed
M$_{\rm h}$, the number of passive galaxies  steadily rises with increasing stellar mass
in analogy to what is observed at $z \simeq$ 0 (cf. for example Lin et al. 2014, McGee et al. 2011).
\par
Similar trends are observed at $z \simeq$ 1; the fraction of star-forming galaxies is a factor
of $\sim$2 higher in the field than in groups at any galaxy stellar mass. Moreover, the 
fractional number of star-forming galaxies within groups is seen to increase with distance from
the group centre, most likely a result of ram-pressure stripping being more efficient in
the central region of groups/clusters where the ICM is densest (Muzzin et al. 2012). In particular,
Muzzin et al. (2014) analysed the velocity-position distribution of post-starburst satellites in 
several 
$z \sim$ 1 clusters, and found that simulations can reproduce it when satellites, upon accretion,
are let to form stars for about 1 Gyr (the time needed to make their first passage at R $\sim$
0.5~R$_{\rm vir}$ where environmental effects become more efficient), and then quench their
star formation activity rapidly, over a time scale between 0.1 and 0.5 Gyr (cf. also Balogh et al. 
2011, Mok et al. 2014).
\par
Our knowledge of galaxy environment at $z \sim$ 1.5 - 2 is clearly much sparser than at $z <$ 1,
since the detection of galaxy clusters at infrared wavelengths is challenging from the ground and
no extensive, imaging and spectroscopic survey has been performed yet at these redshifts that can
give a statistically-robust (against cosmic variance) overview of galaxy properties as a function 
of their haloes. A number of galaxy clusters have been discovered and studied 
beyond $z \sim$ 1.5, which exhibit nearly the same fraction of passive and star-forming satellites,
at odds with their $z <$ 1 counterparts mostly dominated by quenched galaxies. The cluster star-forming
satellites experience star formation rates that are comparable to those measured in the field 
(Hayashi et al. 2010). They are seen to populate the cluster core and outskirts either in equal 
percentage (cf. Hayashi et al. 2010) or to become more numerous in the higher density regions
of these clusters (Tran et al. 2010, Santos et al. 2015). Nevertheless, some clusters at $z >$ 1.5
already exhibit a fraction of passive satellites 
higher than in the field and increasing with decreasing cluster-centric distance, possibly
due to a more efficient ram-pressure stripping in the cluster core (see for example Strazzullo
et al. 2013). The picture thus emerging from these results is one where galaxy clusters at
$z >$ 1.5 are building up their core with the progenitors of the quenched massive galaxies observed
at $z <$ 1.

\section{Conclusions}
The observed properties of satellite galaxies at $z \simeq$ 0 can be ascribed to a star formation
activity that proceeds for 2 - 4 Gyr since accretion onto a bigger halo, and subsequently declines
with an e-folding time $<$ 1 Gyr. These time scales do not significantly depend on halo mass,
and are estimated to be shorter at $z \simeq$ 1: $\sim$1 Gyr and $\sim$0.3 Gyr, respectively. 
How can we reconcile these two snapshots in the time evolution of cosmic structures?
\par
Tinker \& Wetzel (2010) performed clustering measurements for star-forming and passive galaxies
(so distinguished on the basis of their colours) in three publicly available surveys (UKIDSS-UDS, DEEP2 
and COMBO-17), 
and used the fraction of passive satellites to estimate, by comparison with simulations, the satellites 
quenching time and its dependence on redshift. In this case, the quenching time is comparable with
the above mentioned duration of the satellite star formation activity after infall plus the time scale 
of its subsequent rapid quenching. The authors found that the satellite quenching time scale increases
with decreasing redshift as (1 $+ z)^{-1.5}$, in a similar way as the dynamical time scale of dark matter haloes.
This likeness would then suggest that satellites lose their hot gas (aka strangulation) 
mainly because of tidal forces induced by their host halo when they cross it.
This, in concurrence with the ram pressure stripping
of cold/ionized gas, would cause the suppression of star formation in satellites.
\par
An alternative explanation of the redshift dependence of the satellites quenching time comes from
McGee et al. (2014), who analysed the quenching times measured in the literature for satellites at
different redshifts with a model for the baryon cycle in galaxies (i.e. outflows driven by star
formation). The authors suggested that recurring outflows may be able to suppress the star formation activity
of satellites more rapidly than orbit-based gas stripping, especially for the more massive galaxies and
at high redshifts. Dedicated measurements of outflows, their velocities and mass-loading factors, in
central and satellite galaxies at high redshifts will provide an observational assessment of their contribution 
to satellites quenching.   
\par
While satellite galaxies fade and get disrupted, centrals follow a different evolution. As shown
by Pasquali et al. (2009), central galaxies are the place ``most friendly'' to star formation, where star
formation can continue over long time scales. This is because: {\it i)} their environment does not deprive
them of their gas since they do not orbit in their host halo;
{\it ii)} they can accrete gas from their environment, via for example mergers or close
encounters with gas-rich satellites, events that become progressively less likely in more massive haloes given
their higher velocity dispersion. {\it iii)} They might accrete gas from their hot ICM and cool it so 
to trigger a new episode of star formation.
Recently, La Barbera et al. (2014) showed that
central, early-type galaxies in groups (log[M$_{\rm h}$/(h$^{-1}$M$_{\odot})] >$ 12.5) are characterised 
by younger ages, lower [$\alpha$/Fe] ratioes and higher intrinsic reddening than central early-types of the
same velocity dispersion ($\sigma$) but in isolation. Such findings imply that central early-types in groups 
assembled their stellar mass via gas accretion over longer time scales (from $\sim$0.4 Gyr at high $\sigma$ 
to few Gyrs at low $\sigma$) than isolated early-type galaxies.

\acknowledgements
I would like to thank F.C. van den Bosch and A. Gallazzi for their valuable contribution and 
support. I would also like to acknowledge the visitor programme of ESO Vitacura, where part of this 
paper was written.


\end{document}